\documentclass[twocolumn,floatfix,pra,superscriptaddress]{revtex4}
\usepackage{graphicx}
\usepackage{color}
\usepackage{amsmath}

\begin{document}

\title[a]{A Waveguide Frequency Converter Connecting Rubidium Based Quantum Memories to the Telecom C-Band}

\author{Boris Albrecht}
\affiliation{ICFO-The Institute of Photonic Sciences,
Mediterranean Technology park, 08860 Castelldefels, Spain}
\author{Pau Farrera}
\affiliation{ICFO-The Institute of Photonic Sciences,
Mediterranean Technology park, 08860 Castelldefels, Spain}
\author{Xavier Fernandez-Gonzalvo}
\affiliation{ICFO-The Institute of Photonic Sciences,
Mediterranean Technology park, 08860 Castelldefels, Spain}
\affiliation{Current address: Department of Physics, University of
Otago, 730 Cumberland Street, 9016 Dunedin, New Zealand }
\author{Matteo Cristiani}
\affiliation{ICFO-The Institute of Photonic Sciences,
Mediterranean Technology park, 08860 Castelldefels, Spain}
\author{Hugues de Riedmatten}
\affiliation{ICFO-The Institute of Photonic Sciences,
Mediterranean Technology park, 08860 Castelldefels, Spain}
\affiliation{ICREA-Instituci\'o Catalana de Recerca i Estudis
Avan\c cats, 08015 Barcelona, Spain}

\begin{abstract}
The ability to coherently convert the frequency and temporal
waveform of single and entangled photons will be crucial to
interconnect the various elements of future quantum information
networks. Of particular importance in this context is the quantum
frequency conversion of photons emitted by material systems able
to store quantum information, so called quantum memories. There
have been significant efforts to implement quantum frequency
conversion using non linear crystals, with non classical light
from broadband photon pair sources and solid state emitters.
However, so far, solid state quantum frequency conversion has not
been achieved with long lived optical quantum memories. Here, we
demonstrate an ultra low noise solid state photonic quantum
interface suitable for connecting quantum memories based on atomic
ensembles to the telecommunication fiber network. The interface is
based on an integrated waveguide non linear device. As a proof of
principle, we convert heralded single photons at
$780\,\mathrm{nm}$ emitted by a rubidium based quantum memory to
the telecommunication wavelength of $1552\,\mathrm{nm}$. We
demonstrate a significant amount of non classical correlations
between the converted photon and the heralding signal.
\end{abstract}

\maketitle

\section{Introduction}Photonic quantum memories (QM) have been implemented
with single atoms and ions \cite{Kimble2008, Hammerer2010,
Sangouard2011, Ritter2012, Moehring2007, Hofmann2012}, atomic
ensembles \cite{Chaneliere2005, Eisaman2005, Radnaev2010,
Hosseini2011, Bao2012} and solid state systems
\cite{Riedmatten2008, Hedges2010, Clausen2011, Saglamyurek2011}.
Proofs of principle of short distance elementary quantum networks
have been built with these QMs \cite{Chou2007, Yuan2008,
Ritter2012, Moehring2007, Usmani2012, Hofmann2012}. For long
distance implementation of quantum information networks using
quantum repeaters \cite{Kimble2008, Briegel1998, Duan2001,
Sangouard2011}, it is required that optical QMs are connected to
the optical fiber network. However, despite initial attempts to
build QMs operating in the telecom range \cite{Lauritzen2010}, the
best current QMs absorb and emit photons in the visible to near
infrared range, where losses in optical fibers are significant. In
order to overcome this problem, a possible strategy consists in
translating the frequency of the single photons emitted by the QMs
to the telecommunication band, in an efficient, noise free and
coherent fashion.

Quantum frequency conversion (QFC) of photons from rubidium atoms
based QMs to $1367\,\mathrm{nm}$ has been demonstrated using four
wave mixing in a cold and very dense atomic ensemble
\cite{Radnaev2010}, where the input and target wavelengths are
fixed by the available atomic transitions. In contrast, QFC in
solid state non linear materials offers wavelength flexibility and
much simpler experimental implementation, as well as prospects for
integrated circuits using waveguide configurations
\cite{Tanzilli2005, Rakher2010, Ates2012, Takesue2010, Curtz2010,
Zaske2011, Zaske2012, Ikuta2011, DeGreve2012}. Frequency down
conversion using difference frequency generation (DFG) enables
frequency translation from visible to telecom bands and is
therefore ideally suited for quantum repeaters applications
\cite{Sangouard2011}. DFG has been demonstrated with single
photons from solid state emitters \cite{Zaske2012, DeGreve2012}
and spontaneous down conversion sources (SPDC) \cite{Ikuta2011}.
However, so far QFC using solid state non linear devices has not
been demonstrated with long lived optical QMs based on atomic
systems. A significant experimental challenge in solid state QFC
is to reduce the noise generated by the strong pump beam in the
crystal, which is proportional to the input photon duration, in
order to operate in the quantum regime. Now, atomic QMs usually
emit photons with durations ranging from tens of $\mathrm{ns}$
\cite{Felinto2006, Bao2012} to hundreds of $\mathrm{ns}$
\cite{Ritter2012}. This is $1$ to $3$ orders of magnitude longer
than the photons that have been converted so far from SPDC
\cite{Tanzilli2005, Ikuta2011} or broad band solid state emitters
\cite{Zaske2012, DeGreve2012}, which typically have sub
$\mathrm{ns}$ durations. This in turn leads to much stronger
requirements in terms of noise suppression for the QFC.

\begin{figure*}[hbt] \centering
\includegraphics[width=0.7\textwidth]{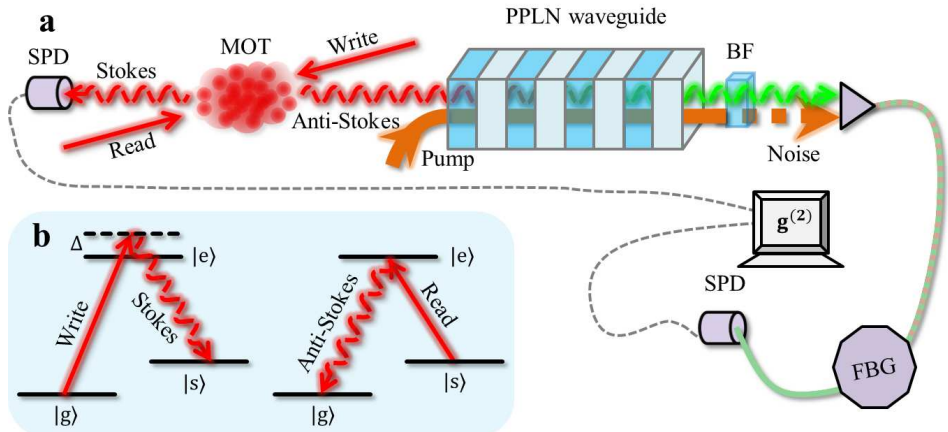}
\caption{{\bf Schematic view of the experimental setup.} {\bf a},
A cold cloud of ${}^{87}\mathrm{Rb}$ atoms confined in a
magneto-optical trap (MOT) serves as quantum memory for light.
Inelastic scattering of a classical {\it write} pulse results in
the emission of a Stokes photon, which heralds the presence of a
collective spin excitation. A subsequent {\it read} pulse maps the
atomic state onto an anti-Stokes photon. The retrieved light is
sent to a PPLN waveguide, together with a strong pump at
$1569\,\mathrm{nm}$. The $780\,\mathrm{nm}$ photons are converted
to $1552\,\mathrm{nm}$ by DFG. After conversion, residual pump
light is blocked by means of a bandpass filter (BF). The converted
light is then coupled in a single mode optical fiber. A fiber
Bragg grating filters out the broadband noise generated by the
pump. The non classical nature of the storage, retrieval and
conversion processes is probed by measuring the cross-correlation
between Stokes and (un)converted anti-Stokes photons. {\bf b},
Level scheme used for the DLCZ memory. The {\it write} pulse
off-resonantly couples the ground state $\left| g \right\rangle $
to the excited level $\left| e \right\rangle $. A Stokes photon is
emitted with probability $p$, thus creating a single collective
excitation to the storage state $\left| s \right\rangle $. During
the retrieval process, a pulse resonant with the $\left| s
\right\rangle \rightarrow \left| e \right\rangle$ transition ({\it
read} beam) collectively transfers back the atom to the initial
state $\left| g \right\rangle$.}\label{Figure1}
\end{figure*}

In this paper we demonstrate an ultra low noise solid state
photonic quantum interface capable of connecting QMs based on
atomic ensembles to the telecommunication network. We generate
heralded single photons at $780\,\mathrm{nm}$ from a cold
$\mathrm{{}^{87}Rb}$ atomic ensemble QM and convert them to
$1552\,\mathrm{nm}$ using frequency down conversion in a non
linear periodically poled lithium niobate (PPLN) waveguide. By
combining high QM retrieval efficiency, high QFC efficiency and
narrow band filtering, we can operate the combined systems in a
regime where a significant amount of non classical correlations is
preserved between the heralding and converted photons.

\begin{figure*}[hbt]
\centering
\includegraphics[width=0.75\textwidth]{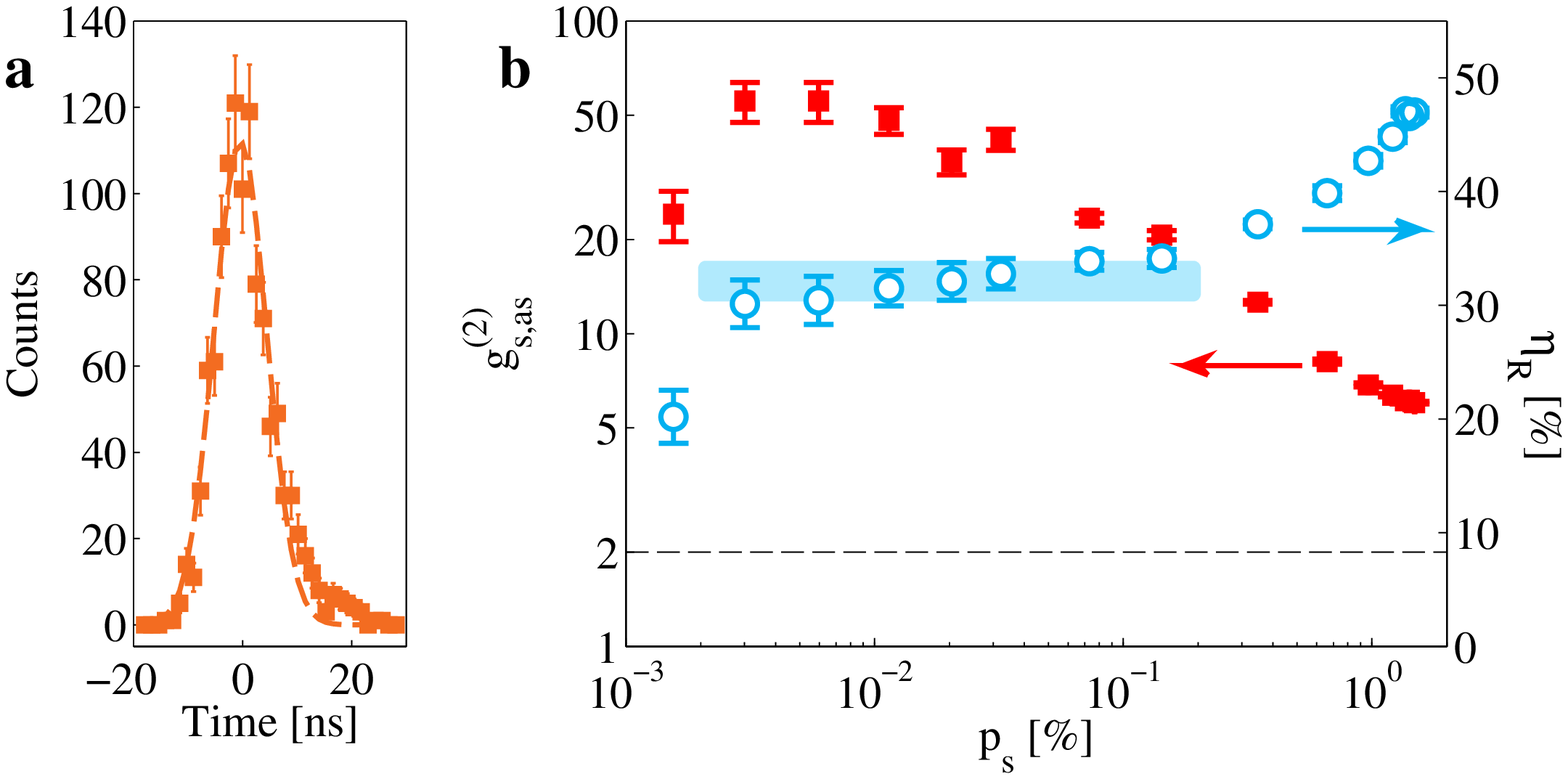}
\caption{{\bf Characterization of the DLCZ quantum memory.} {\bf
a}, Temporal shape of anti-Stokes photons. The number of
anti-Stokes detection events per time bin (bin size
$1.28\,\mathrm{ns}$) conditioned on a Stokes detection over
$5\times10^6$ trials is plotted as a function of the arrival time
(orange squares). The data shown correspond to $p_{s} = 0.15\,\%$.
The dashed line is a gaussian fit with $\mathrm{FWHM} =
11.4(9)\,\mathrm{ns}$. For this data, we have $g^{(2)}_{s,
as}=22(1)$ and $\eta_R =3 5(1)\,\%$. {\bf b} Second-order
cross-correlation function $g^{(2)}_{s, as}$ (red plain squares,
left axis) and retrieval efficiency $\eta_R$ (blue empty circles,
right axis) as a function of the Stokes detection probability
$p_{s}$. For $p_{s}$ between $0.002$ and $0.2\,\%$ (shaded area),
$\eta_R$ is constant ($32(2)\%$). The $g^{(2)}_{s, as}$ function
increases until $56(8)$, before dropping for lower values of $p_s$
due to detector dark counts. The dashed line is the
$g^{(2)}_{s,as}$ classical limit for two mode squeezed states.}
\label{Figure2}
\end{figure*}

\section{Results}
\subsection{The atomic quantum memory}Our experimental setup is
shown in Fig. \ref{Figure1} (see Appendix \ref{app:A} for
details). It is composed of two main parts, the QM and the quantum
frequency conversion device (QFCD). The QM is implemented with an
ensemble of laser cooled $\mathrm{{}^{87}Rb}$ atoms, following the
scheme proposed by Duan, Lukin, Cirac and Zoller \cite{Duan2001}.
Single collective spin excitations are created by a
$16\,\mathrm{ns}$ long weak write pulse at $780\,\mathrm{nm}$,
blue detuned by $40\,\mathrm{MHz}$ from the $\left|g\right\rangle$
to $\left|e\right\rangle$ transition, and heralded by the
detection of a photon emitted by spontaneous Raman scattering,
called Stokes photon. The collective spin excitation is stored in
the atoms for a programmable time ($330\,\mathrm{ns}$ in our
experiment) before being transferred back into a single photon
(anti-Stokes) thanks to an $11\,\mathrm{ns}$ long read laser pulse
resonant with the $\left|s\right\rangle$ to $\left|e\right\rangle$
transition. The anti-Stokes photon is emitted with high efficiency
in a single spatio-temporal mode thanks to collective interference
between all the emitters. It is then coupled with high efficiency
($0.7$) into a single mode fiber before being detected by a single
photon detector (SPD) with efficiency $\eta_{d,780}=0.43$.

Figure \ref{Figure2}a shows the temporal profile of the heralded
retrieved anti-Stokes field. We measure a full width at half
maximum ($\mathrm{FWHM}$) of $11.4(9)\mathrm{ns}$, shorter than
previous single photons generated by the same method
\cite{Felinto2006, Bao2012}. We then characterize the non
classical correlations between the Stokes photons and the stored
spin excitations. This is realized by measuring the second order
cross-correlation function $g^{(2)}_{s,as}$ between  the Stokes
and anti-Stokes fields, defined as $g^{(2)}_{s,as} = p_{s,as} /
(p_{s} p_{as})$, where $p_{s,as}$ is the probability to detect a
coincidence between the two modes and $p_{s}$ ($p_{as}$) is the
probability to detect a Stokes (anti-Stokes) photon. $p_{s}$ is
proportional to the write pulse intensity. Figure \ref{Figure2}b
shows a typical measurement of $g^{(2)}_{s,as}$ as a function of
$p_s$. As expected for this type of source, the degree of
correlation increases when decreasing the excitation probability
\cite{Laurat2006}. The measured $g^{(2)}_{s,as}$ is well above the
classical threshold of $2$ for two-mode squeezed states (see
Appendix \ref{app:D}). This suggests that strong non classical
correlations between the Stokes photon and the stored spin
excitation are present. Additional measurements demonstrating the
non classical character of the correlations are shown in Appendix
\ref{app:D} (see also Table \ref{tab:g2auto}). Also shown in Fig.
\ref{Figure2}b is the retrieval efficiency $\eta_R =
p_{s,as}/(p_s\eta_{d,780})$ corresponding to the probability to
find an anti-Stokes photon before the SPD conditioned on the
detection of a Stokes photon. In the region where the
multi-excitation probability and the Stokes detector dark counts
are negligible \cite{Laurat2006} ($p_{s}$ between $0.002\,\%$ and
$0.2\,\%$), $\eta_R$ is constant and equal to $32(2)\,\%$ (shaded
area on Fig. \ref{Figure2}b).

\begin{figure*}[hbt] \centering
\includegraphics[width=0.75\textwidth]{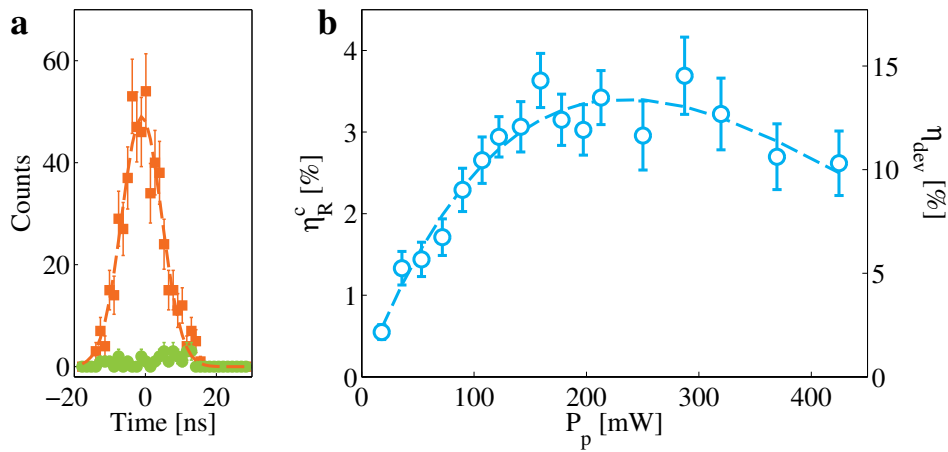}
\caption{{\bf Characterization of the quantum frequency conversion
device} {\bf a}, Converted anti-Stokes photon shape. The
conditioned events are plotted as orange squares, while the
conditioned pump noise is shown as green circles. The measured
signal to noise ratio is $18(5)$. The time bin size is
$1.28\,\mathrm{ns}$. The Stokes detection probability for the
measurement is $p_{s} = 0.16\,\%$ while the pump power is
$P_{p}=120\,\mathrm{mW}$. For this measurement, we have
$g^{(2)c}_{s, as}=11.6(7)$ and $\eta_R^{c}=3.4(2)\,\%$ {\bf b}
Conversion efficiency as a function of pump power after the
waveguide measured with heralded anti-Stokes photons. The right
axis shows the device efficiency $\eta_{\mathrm{dev}}$
corresponding to raw conversion efficiency corrected only by the
detection efficiency ($0.1$). The left axis shows
$\eta_R^{c}=\eta_R^{in}\eta_{\mathrm{dev}}$, i.e. the probability
to find a converted photon before the SPD, conditioned on a Stokes
detection. The Stokes detection probability for this measurement
is $p_{s} = 0.16\,\%$.} \label{Figure3}
\end{figure*}

\subsection{The Quantum Frequency Converter Device}The heralded
anti-Stokes single photon is then directed towards the QFCD where
the conversion from $780\,\mathrm{nm}$ to $1552\,\mathrm{nm}$
takes place. From the measured $\eta_R$ and the optical
transmission between the QM and the QFCD
($\eta_{\mathrm{loss}}=0.77$), we infer that, conditioned on a
Stokes detection, the number of photons before the QFCD is
$\eta^{in}_{R}=\eta_R\eta_{\mathrm{loss}}=0.25(2)$. The pump light
at $1569\,\mathrm{nm}$ and the single photons at
$780\,\mathrm{nm}$ are combined using a dichroic mirror. Both
beams are then coupled into the waveguide where the conversion
takes place. After conversion, the pump light is blocked by means
of two bandpass filters, and the converted light is coupled into a
single mode optical fiber. A fiber Bragg grating with a bandwidth
of $2.5\,\mathrm{GHz}$ and a transmission efficiency of $0.7$ then
filters out the broadband noise generated in the crystal by the
pump beam \cite{Fernandez-Gonzalvo2013}. The single photons at
$1552\,\mathrm{nm}$ are finally detected with an InGaAs avalanche
photodiode SPD (detection efficiency $\eta_{d,1552}=0.1$).

Figure \ref{Figure3}a shows the waveform of the heralded single
photons after conversion. The measured temporal profile
($\mathrm{FWHM} = 13(1)\,\mathrm{ns}$) is very similar to the
input one, which shows that the conversion preserves the waveform
of the photons. The noise generated by the pump beam at
$1569\,\mathrm{nm}$ is measured by blocking the input of the
converter. The measured signal to noise ratio ($\mathrm{SNR}$) of
$18(5)$ gives an upper bound for the value of the cross
correlation function $g^{(2)c}_{s,as}$ achievable after conversion
(see Appendix \ref{app:C}). We also measure the efficiency of the
conversion process as a function of the pump power $P_{p}$
measured after the waveguide. Figure \ref{Figure3}b shows the
device efficiency $\eta_{\mathrm{dev}} = p_{s,as}^{c} /( p_s
\eta_R^{in} \eta_{d,1552})$, where $p_{s,as}^{c}$ is the
probability to detect a coincidence after conversion.
$\eta_{\mathrm{dev}}$ corresponds to the probability to find a
converted photon in a single mode optical fiber after the QFCD
(including filtering and fiber coupling) for a single photon
input. The data are fitted with the following formula
$\eta_{\mathrm{dev}}=\eta_{\mathrm{max}}\sin^2
(L\sqrt{\eta_{n}P_{p}})$ \cite{Zaske2011}. From the fit, we
extract $\eta_{n}=120(10)\,\mathrm{\% W^{-1} cm^{-2}}$ and
$\eta_{\mathrm{max}}=13.6(12)\,\%$. The maximum achievable device
efficiency is limited in our current setup to $17.6\,\%$ by
optical losses, including fiber coupling efficiency($60\,\%$) and
Bragg grating filter transmission ($70\,\%$) at the converted
wavelength of $1552\,\mathrm{nm}$, as well as by the waveguide
incoupling ($60\,\%$) and transmission ($70\,\%$) efficiencies for
$780\,\mathrm{nm}$ photons. The discrepancy between the measured
$\eta_{\mathrm{max}}$ and the theoretical one is attributed to non
perfect mode overlap between the pump and single photon in the
waveguide \cite{Zaske2011}. Figure \ref{Figure3}b also displays
$\eta_R^{c}=\eta_R^{in}\eta_{\mathrm{dev}}$, which can be
considered as the combined efficiency of the QM+QFCD system,
including all losses.

\begin{figure}[hbt]
\centering
\includegraphics[width=0.5\textwidth]{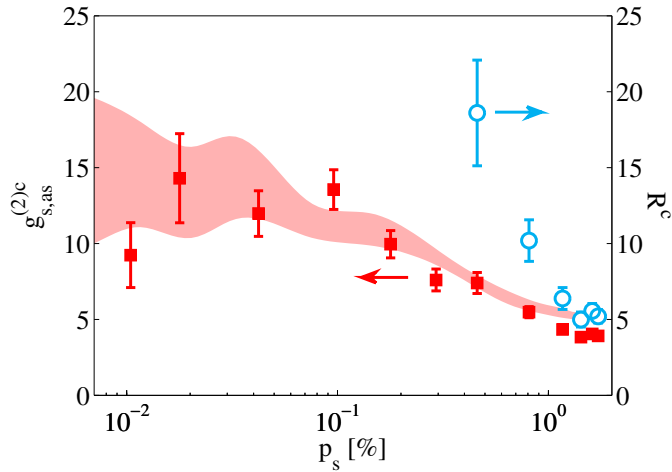}
\caption{{\bf Characterization of the quantum memory--quantum
frequency converter combined setup.} Cross-correlation between
Stokes and converted anti-Stokes photons as a function of Stokes
detection probability, for a pump power of $140\,\mathrm{mW}$ (red
plain squares, left axis). The shaded area represents the expected
$g^{(2)c}_{s,as}$ deduced from the data of Fig. \ref{Figure2}b by
means of equation (\ref{equation1}) (see Appendix \ref{app:B}).
The measured Cauchy-Schwartz parameter $R^c$ is plotted on the
right axis (blue empty circles). The unconditional autocorrelation
measurement for the converted anti-Stokes field has been measured
for a pump power of $120\,\mathrm{mW}$.} \label{Figure4}
\end{figure}

\subsection{Quantum state preservation}In order to verify that the
conversion preserves the quantum character of the input light, we
measure the second order cross-correlation function after
conversion $g^{(2)c}_{s,as}$ as a function of $p_s$ (see Fig.
\ref{Figure4}). As in the case before conversion (Fig.
\ref{Figure2}b), we also observe an increase of non classical
correlations when decreasing the excitation probability, but the
maximum is now reached for much higher values of $p_s$ because of
the background noise induced by the converter pump laser. However,
this background is sufficiently low to observe strong non
classical correlations between the heralding photons at
$780\,\mathrm{nm}$ and the converted photons at
$1552\,\mathrm{nm}$. The expected value of the $g^{(2)c}_{s,as}$
after conversion can be estimated using a simple model taking into
account the noise induced by the pump laser, using the following
expression (see Appendix \ref{app:B}):

\begin{equation}
\label{equation1} g^{(2)c}_{s,as}=g^{(2)}_{s,as}
\frac{\mathrm{SNR} + 1}{\mathrm{SNR} + g^{(2)}_{s,as}}
\end{equation}

The shaded area displayed in Fig. \ref{Figure4} corresponds to the
expected values taking into account the $g^{(2)}_{s,as}$ measured
in Fig. \ref{Figure2} and $\mathrm{SNR}=\eta_{R}^{in}\cdot
\mathrm{SNR}^\mathrm{max}$, where $\mathrm{SNR}^\mathrm{max}$ is
the maximal achievable $\mathrm{SNR}$ for $\eta_R^{in}=1$ and can
be accurately measured with weak coherent states (see Appendix
\ref{app:B}). We find $\mathrm{SNR}^\mathrm{max}=85(3)$. The good
agreement between experimental data and expected values suggests
that the noise generated from the pump is the only factor
degrading the correlations and that further reduction of this
noise should lead to even higher non classical correlations.

Finally, in order to unambiguously prove the non classical
character of the correlations, we also measure the unconditional
auto correlation function for the Stokes ($g^{(2)}_{s,s} = p_{s,s}
/ (p_{s} p_{s}))$ and converted anti-Stokes fields
($g^{(2)c}_{as,as} = p_{as,as}^{c} / (p_{as}^{c} p_{as}^{c}))$.
For classical independent fields, the Cauchy-Schwartz inequality
$R^c=\left(g^{(2)c}_{s,as}\right)^2/ \left(g^{(2)}_{s,s} \cdot
g^{(2)c}_{as,as}\right) \leq 1$ must be satisfied. The values for
$g^{(2)}_{s,s}$ and $g^{(2)c}_{as,as}$ are listed in Table
\ref{tab:g2auto}. On the right axis of Fig. \ref{Figure4} we plot
the values of $R^c$ measured in our experiment for various $p_s$.
The Cauchy-Schwartz inequality is clearly violated, which is a
proof of non classical correlation between Stokes and converted
anti-Stokes photons.

\section{Discussion}We now discuss the performances of our interface. The
maximal $g^{(2)c}_{s,as}$ obtained is about $15$, which would be
sufficient to enable a violation of Bell's inequality if the
photons were entangled \cite{Riedmatten2006}. The measured maximal
$\eta_{\mathrm{dev}}=13.6(12)\,\%$ corresponds to an internal
waveguide conversion efficiency of $77\,\%$. As it is, it would
already lead to an increased transmission with respect to the
unconverted photon at $780\,\mathrm{nm}$ ($3\,\mathrm{dB/km}$
loss) after only $\sim 3\,\mathrm{km}$ of fiber (see also Fig.
\ref{fig:S3}). Further improvement in the device efficiency could
be primarily obtained by technical improvements such as decreasing
the passive optical losses due to waveguide coupling and
transmission, fiber coupling and filtering transmission.

Our quantum photonic interface could be directly useful for some
efficient quantum repeater architectures using DLCZ type QMs,
where entanglement between remote quantum nodes is achieved by
entanglement swapping involving the anti-Stokes photons
\cite{Sangouard2008a}. In that case, the anti-Stokes photons are
sent over long distances and must thus be converted to telecom
wavelengths. For protocols where entanglement between remote
atomic ensembles is achieved by detection of the Stokes photon,
such as the original DLCZ protocol \cite{Duan2001}, or in order to
achieve entanglement between a telecom photon and a stored spin
excitation, the Stokes photon should then be converted. This would
however require a significant increase of the $\mathrm{SNR}$
compared to our present experiment, since the Stokes emission
probability has to be low to allow highly non classical
correlations. This reduction may be achieved by using narrower
filtering  since the current value of $2.5\,\mathrm{GHz}$ is about
two order of magnitude larger than the photon bandwidth. Obtaining
higher $\mathrm{SNR}$ would be also interesting in view of using
this interface in combination with single atom QMs emitting
photons with duration of a few hundred $\mathrm{ns}$
\cite{Ritter2012}.

We have demonstrated quantum frequency conversion to
telecommunication wavelength of single photons emitted by a
quantum memory, using a solid state integrated photonic quantum
interface. These results open the road to the use of integrated
optics as a practical and flexible interface capable of connecting
QMs to the optical fiber network. The wavelength flexibility
offered by non linear crystals opens the door to QFC with other
kind of QMs, such as solid state QMs \cite{Riedmatten2008,
Hedges2010, Clausen2011, Saglamyurek2011} or single ions
\cite{Moehring2007}. Our work is a step to extend quantum
information networks to long distances and opens prospects for the
coupling of different types of remote QMs via the optical fiber
network.

\appendix \section{Atomic ensemble based quantum memory}
\label{app:A}Our quantum memory (QM) is based on an ensemble of
rubidium atoms, trapped and cooled in a magneto optical trap (MOT)
operating on the $\mathrm{D2}$ line of ${}^{87}\mathrm{Rb}$ at
$780\,\mathrm{nm}$. The optical depth (OD) of the sample is
measured to be $\sim 12$, when probing the $\left|5\,{}^2S_{1/2},
F=1\right\rangle \rightarrow \left|5\,{}^2P_{3/2},
F^{\prime}=2\right\rangle$ transition. At the beginning of the
experimental sequence the MOT is turned off and the atoms are
optically pumped during $300\,\mathrm{\mu s}$ into the ground
state $\left|g\right\rangle = \left|5\,{}^2S_{1/2},
F=1\right\rangle$. The initial state of the system is described by
$\left|G\right\rangle = \left|g_1 \dots g_N\right\rangle$, $N$
being the atom number. A weak laser pulse ({\it write} beam),
linearly polarized and blue detuned by $40\,\mathrm{MHz}$ with
respect to the $\left|g\right\rangle \rightarrow
\left|e\right\rangle$ transition ($\left|e\right\rangle$ being the
excited state $\left|5\,{}^2P_{3/2}, F^{\prime}=2\right\rangle$),
is then sent to the cold cloud. The {\it write} beam has a waist
of $\sim150\,\mathrm{\mu m}$ and the pulse is characterized by a
full width at half maximum ($\mathrm{FWHM}$) of $16\,\mathrm{ns}$.
Raman scattering of the {\it write} beam induces the transfer of
one atom to the storage level $\left|s\right\rangle =
\left|5\,{}^2S_{1/2}, F=2\right\rangle$, with low probability.
This process is followed by the emission of a photon on the
$\left|e\right\rangle \rightarrow \left|s\right\rangle$ transition
(Stokes photon), which is collected by means of a
$500\,\mathrm{mm}$ focal length lens and then coupled into a
single mode optical fiber. The collection mode corresponds to a
waist of $\sim50\,\mathrm{\mu m}$, such that the observed region
is smaller than the volume illuminated by the {\it write} beam. We
indicate with $p$ the probability for a Stokes photon to be
emitted in the collection mode. The emitted light is finally
detected by a Si avalanche photodiode (Count-100C-FC, Laser
components), characterized by a measured detection efficiency of
$\sim43\%$ at $780\,\mathrm{nm}$ and a dark count rate of
$100\,\mathrm{Hz}$. The single photon detector (SPD) is operated
in gated mode with a detection window of $40\,\mathrm{ns}$.

\begin{figure*}[hbt!]
\centering
\includegraphics[width=0.95\textwidth]{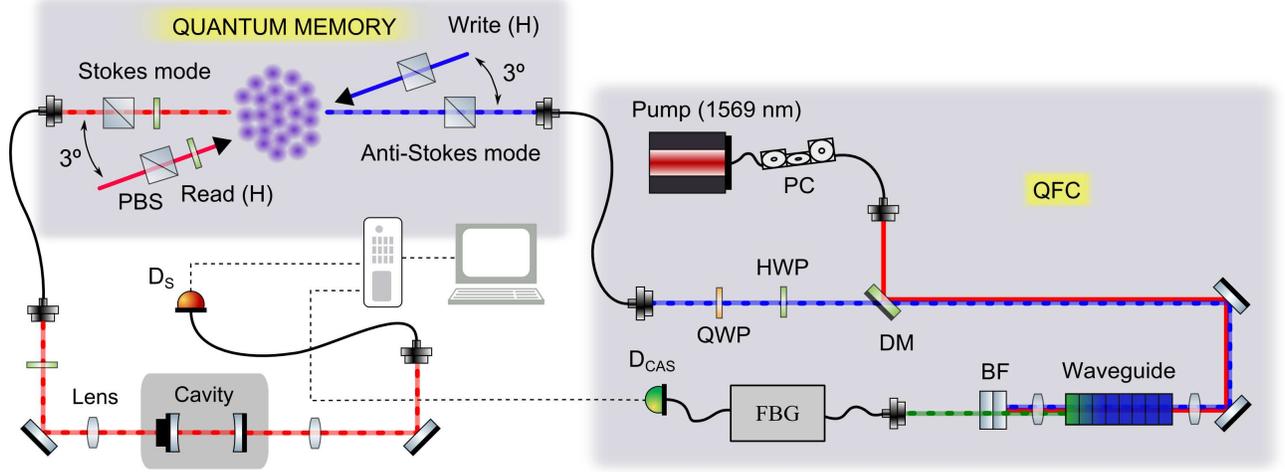}
\caption{{\bf Schematic view of the experimental setup.} See text
for a detailed description of the different components.}
\label{fig:S1}
\end{figure*}

Since the Stokes photons are emitted while the {\it write} pulse
is on, it is necessary to filter the Stokes field from the
possible leakage of the classical beam. For this purpose three
strategies are employed: (a) polarization, (b) angular and (c)
frequency filtering (see Fig. \ref{fig:S1} for a schematic view of
the experimental setup). (a) By means of a half waveplate and a
polarizing beam splitter cube, we ensure to detect only Stokes
photons orthogonally polarized with respect to the {\it write}
beam. (b) The detection mode is aligned with an angle of $3$
degrees with respect to the {\it write} mode, thus avoiding direct
coupling of the classical light into the collection fiber. (c) In
between the atomic cloud and the SPD we place a monolithic
Fabry-Perot cavity characterized by a $\mathrm{FWHM}$ of
$\sim60\,\mathrm{MHz}$ and a free spectral range (FSR) of
$\sim12\,\mathrm{GHz}$. The cavity is formed by a plano-concave
lens coated on both sides with a high reflectivity layer at
$780\,\mathrm{nm}$, following the design described by
Palittapongarnpim et al. \cite{Palittapongarnpim2012}. The Stokes
photons are delivered by an optical fiber to a free-space setup
where they are shaped to match the $\mathrm{TE_{00}}$ mode of the
resonator. The cavity length is tuned by acting on the lens
temperature and is adjusted to be resonant with the Stokes
photons. The cavity transmission is $\sim50\%$ when measured with
monochromatic light. Considering that our Stokes photons have a
spectral bandwidth of about $40\,\mathrm{MHz}$ (i.e.
$\mathrm{FWHM}$ of $11\,\mathrm{ns}$) their transmission through
the resonator is limited to $\sim80\%$. The maximum cavity
transmission is then $\sim40\,\%$. The photons filtered by the
resonator are then coupled into another fiber connected to the
SPD. The total transmission of the system, including losses and
fiber in-coupling efficiency, is measured to be $\sim25\,\%$ when
probed with continuous light (i.e. $20\,\%$ for the Stokes
photons). The {\it write} beam is detuned with respect to the
Stokes field by $6.8\,\mathrm{GHz}$ (i.e. the hyperfine splitting
between the $\left|g\right\rangle$ and $\left|s\right\rangle$
levels) which corresponds to about half the FSR of the cavity.
This ensures that its transmission through the resonator is below
$1\,\%$. The combination of the three filtering methods described
above allows us to reduce the noise caused by the {\it write}
light to a level lower than the dark count rate of the SPD.

The probability to detect a Stokes photon is indicated as $p_s$
and it is given by $\eta_s \times p$, where the factor $\eta_s =
6\,\%$ accounts for the coupling efficiency of the Stokes mode
into the optical fiber ($0.7$, see later for details), the filter
cavity transmission ($\sim0.2$) and the detection efficiency
($0.43$). The Stokes creation probability $p$ can be varied acting
on the {\it write} power. In our experiment $p$ is spanned over
three orders of magnitude from $2.5\times10^{-4}$ to $0.25$ (i.e.
$p_s$ varies from $1.5\times10^{-5}$ to $1.5\times10^{-2}$). The
detection of a Stokes photon projects the system onto the
collective state:
\begin{equation}
\left|S\right\rangle = \frac{1}{\sqrt{N}}
\sum\limits_{j=1}^N\exp{\left[\imath(\vec{k}_w -
\vec{k}_s)\cdot\vec{x}_j\right]}\left|g_1 \dots s_j \dots
g_N\right\rangle,
\end{equation}
where $\vec{x}_j$ is the position of the $j$-th atom, while
$\vec{k}_w$ and $\vec{k}_s$ are the wavevectors for the {\it
write} and Stokes modes, respectively. This is a good description
of the system only under the assumption of low excitation
probability (i.e. $p\ll1$), such that multiple excitations can be
neglected. This process can be regarded as a heralded generation
of a single spin excitation in the atomic ensemble.

After a programmable delay $\tau_s$ ($330\,\mathrm{ns}$ for our
experiment), a bright light pulse ({\it read} beam) resonantly
couples the $\left|s\right\rangle$ and $\left|e\right\rangle$
states, thus transferring back the population to the ground level
$\left|g\right\rangle$. The pulse propagates in the same spatial
mode as the {\it write} beam but in the opposite direction. The
peak power of $\sim800\,\mathrm{\mu W}$ and the $\mathrm{FWHM}$ of
$11\,\mathrm{ns}$ have been chosen in order to maximize the
transfer efficiency \cite{Mendes2013}. The retrieval process is
accompanied by the collective re-emission of a photon on the
$\left|e\right\rangle \rightarrow \left|g\right\rangle$ transition
(anti-Stokes photon). At the end of the sequence the collective
state of the system is:
\begin{widetext}
\begin{equation}
\left|G^\prime\right\rangle = \frac{1}{\sqrt{N}}
\sum\limits_{j=1}^N\exp{\left[\imath(\vec{k}_w -
\vec{k}_s)\cdot\vec{x}_j\right]}\exp{\left[\imath(\vec{k}_{r} -
\vec{k}_{as})\cdot\vec{x}_j^\prime\right]}\left|g_1 \dots g_j
\dots g_N\right\rangle,
\end{equation}
\end{widetext} where $\vec{x}_j^\prime$ is the final position of
the $j$-th atom, while $\vec{k}_r$ and $\vec{k}_{as}$ are the
wavevectors for the {\it read} and anti-Stokes modes respectively.
For a sample temperature of the order of a few hundreds of micro
Kelvin and a storage time below $1\,\mathrm{\mu s}$, the atomic
motion is negligible during the storing process and
$\vec{x}_j^\prime \simeq \vec{x}_j$. In this case, the final state
$\left|G^\prime\right\rangle$ coincides with the initial state
$\left|G\right\rangle$ if the phase matching condition $\vec{k}_w
- \vec{k}_s = \vec{k}_{as} - \vec{k}_r$ is satisfied. For our
excitation geometry this implies that the anti-Stokes photon is
emitted in the same mode as for the Stokes photon but in opposite
direction (i.e. $\vec{k}_{as} = -\vec{k}_s$). The single spin
excitation created in the ensemble during the {\it write} process
is then collectively read-out and mapped onto the anti-Stokes
photon.

The retrieved photons are coupled into an optical fiber. The
overlap between the Stokes and anti-Stokes modes is estimated to
be $\sim70\%$ by measuring the fiber to fiber coupling using
classical light. The photons are then delivered to a second SPD
with similar specifications as the first one (SPCM-AQRH-14-FC,
Excelitas), temporally gated for $40\,\mathrm{ns}$. Polarization
and angular filtering are used to reduce the leakage of the {\it
read} beam as for the Stokes detection. In this case, however, we
do not use a cavity for frequency filtering, in order to maximize
the retrieval efficiency. This is crucial when the system is used
in combination with the quantum frequency converter device (QFCD).

A {\it clean} pulse with the same frequency and spatial mode as
the {\it read} beam is sent to the atomic cloud in order to
efficiently pump all the atoms back to the initial state
$\left|g\right\rangle$, while the SPDs are kept off. The {\it
write--read} sequence is repeated $1000$ times, each trial lasting
$\Delta\tau = 1.4\,\mathrm{\mu s}$. During the total interrogation
time of $1.4\,\mathrm{ms}$ the atomic cloud is falling freely
under the effect of gravity. During this short time the OD
reduction due to cloud expansion is negligible. We can therefore
assume that all the trials are performed under the same
conditions. The trap and repumper beams as well as the magnetic
field gradient are then switched on and the atoms are recaptured
in the MOT, where they are laser cooled during $20\,\mathrm{ms}$
before the sequence starts again. This full cycle is repeated
until a good statistic is reached.

\begin{figure}[hbt!]
\centering
\includegraphics[width=0.5\textwidth]{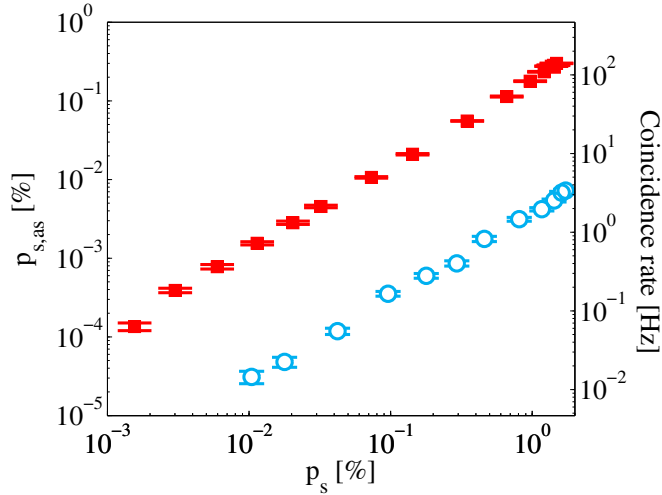}
\caption{{\bf Heralded anti-Stokes generation rates.} The
probability $p_{s,as}$ to detect a Stokes--anti-Stokes coincidence
per trial is plotted as a function of the Stokes detection
probability $p_{s}$ for the memory alone (red plain squares) and
in combination with the QFCD (blue empty circles). On the right
axis the corresponding detection rates are also plotted.}
\label{fig:S2}
\end{figure}

Due to the strong correlations between the Stokes photon and the
single spin excitation created in the system, if we assume that
the read-out process is performed with high efficiency, we can
describe the joint state of the Stokes and anti-Stokes photons as:
\begin{equation}
\label{eq:squeezed}
\begin{split}
\left|\psi\right\rangle = &
\left(1 - \frac{p}{2}\right) \left| 0_s, 0_{as} \right\rangle +
\\
& + \sqrt{p} \left| 1_s, 1_{as} \right\rangle +\\
& + p \left| 2_s, 2_{as} \right\rangle + O\left(p^{3/2}\right).
\end{split}
\end{equation}
This ideal two-mode squeezed state displays a high degree of
second order cross-correlations between the Stokes and anti-Stokes
fields. The second order cross-correlation function
$g^{(2)}_{s,as}$ is defined as $p_{s,as}  / \left( p_s p_{as}
\right)$, where $p_s$ and $p_{as}$ are the probabilities to detect
a Stokes and an anti-Stokes photons per trial, respectively. The
quantity $p_s \cdot p_{as}$ can be interpreted as the probability
to get an accidental coincidence. On the other hand, $p_{s,as}$ is
the probability to detect an anti-Stokes photon conditioned on the
detection of a Stokes photon in the same {\it write-read} trial
(i.e the Stokes--anti-Stokes coincidence probability). In Fig.
\ref{fig:S2}, $p_{s,as}$ is plotted as a function of the Stokes
detection probability $p_s$, together with the coincidence rate
achievable in the experiement. In order to measure the
$g^{(2)}_{s,as}$ function, the outputs of the two SPDs are
acquired by a time-stamping card which records their arrival
times. Correlated photons are emitted within the same {\it
write--read} trial and therefore their relative arrival time is
given by $\tau_s$. On the other hand, Stokes and anti-Stokes
photons generated in different trials will arrive with a time
difference $\tau_s + n \cdot \Delta\tau$, with $n$ integer. The
ratio between the number of coincidences with an arrival time of
$\tau_s$ and $\tau_s+\Delta\tau$ is therefore a measurement of the
second order cross-correlation function.

\section{Quantum frequency converter}
\label{app:B} Our quantum frequency converter is based on
difference frequency generation (DFG) of $1552\,\mathrm{nm}$
light, achieved by combining input photons at $780\,\mathrm{nm}$
with a strong pump at $1569\,\mathrm{nm}$ in a periodically poled
lithium niobate (PPLN) waveguide. Our current experimental setup
(see Fig. \ref{fig:S1}) is based on the apparatus described in
\cite{Fernandez-Gonzalvo2013}. Thanks to the introduction of a new
filtering stage the noise contribution has been reduced by more
than one order of magnitude with respect to our previous work.
This achievement, together with an improvement of the device
efficiency and the use of shorter photons, allowed us to increase
the signal to noise ratio of the conversion process by almost two
orders of magnitude (see later for details). In the following,
after a brief description of the main parts of the setup, we will
focus on the characteristics of the new spectral filter,
suggesting the reader to refer to our previous work for an
exhaustive description of the other elements.

\begin{figure}[hbt!]
\centering
\includegraphics[width=0.5\textwidth]{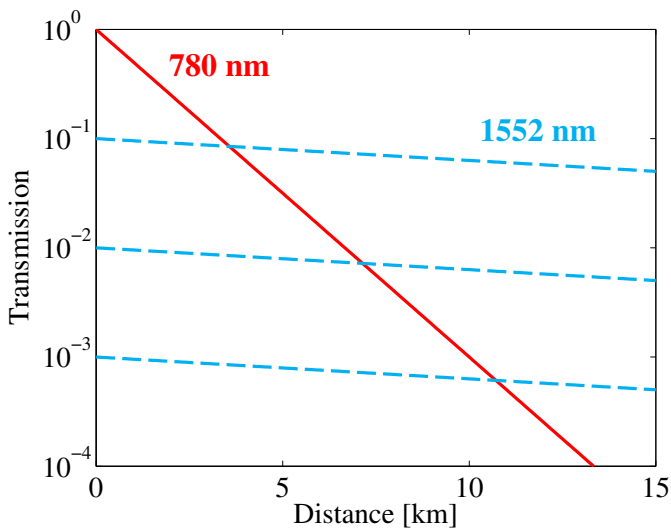}
\caption{{\bf Comparison between transmissions of converted and
unconverted photons.} Transmission of $780\,\mathrm{nm}$ light in
a telecom fiber as a function of the propagation distance assuming
losses of $\sim3\,\mathrm{dB/km}$ (red continuous line).
Similarly, the blue dashed lines represent transmission of
$1552\,\mathrm{nm}$ converted photons with $10$, $1$ and $0.1\,\%$
device efficiency, assuming propagation losses of
$\sim0.2\,\mathrm{dB/km}$. As it can be seen, even for a device
efficiency as low as $0.1\,\%$, the limited conversion efficiency
is compensated after only $11\,\mathrm{km}$ of propagation through
the fiber.} \label{fig:S3}
\end{figure}

The input photons at $780\,\mathrm{nm}$ are combined on a dichroic
beam splitter together with the pump light, obtained from an
erbium doped fiber amplifier feeded by an external cavity diode
laser. The two light fields are then coupled to the nonlinear
waveguide by means of an aspheric lens. The measured coupling
efficiencies are $68\%$ and $60\%$ for the pump and input modes,
respectively. The input lens has a transmission of $\sim66\,\%$ at
$1569\,\mathrm{nm}$. The specified waveguide transmissions are
about $62\,\%$ and $78\,\%$ at the input and pump wavelengths,
respectively. At the waveguide output, two band pass filters with
a $\mathrm{FWHM}$ of $7\,\mathrm{nm}$ isolate the converted signal
from the residual pump light. Each filter has a transmission of
$93\,\%$ at $1552\,\mathrm{nm}$, while it displays an OD of
$\sim12$ at $1569\,\mathrm{nm}$. This filtering stage is not
sufficient to reduce the noise around the converted wavelength of
$1552\,\mathrm{nm}$ below the single photon level. The strong pump
light generates additional noise whose broadband nature has been
investigated in \cite{Fernandez-Gonzalvo2013}. In particular, we
measured that the residual noise level is proportional to the
spectral bandwidth of the filtering stage (see also
\cite{kuo2013}).

Guided by this observation, we replaced our previous filtering
stage, based on a simple diffraction grating, with a fiber Bragg
grating. The current filter linewidth of $2.5\,\mathrm{GHz}$ is
$34$ times narrower than the minimum we could achieve with the
previous setup. When the QM is used in combination with the QFCD,
the fiber Bragg grating helps filtering the residual leakage of
the {\it read} beam, like the monolithic resonator used for the
Stokes photon. Considering the coupling efficiency of the input
light into the waveguide ($60\,\%$), the intrinsic losses in the
waveguide ($70\,\%$, corresponding to the geometric average of the
losses experienced at $780\,\mathrm{nm}$ and $1552\,\mathrm{nm}$),
the fiber coupling efficiency ($60\%$) and the transmission of the
filtering elements ($70\%$), the achievable total efficiency is
limited to $17.6\,\%$. The converted light is finally sent to an
InGaAs SPD (id220, Id Quantique), with a detection efficiency of
$10\%$ and a dark count rate of $400\,\mathrm{Hz}$ with a dead
time of $20\,\mathrm{\mu s}$.

\subsection{QFCD characterization}
\begin{figure*}[hbt]
\centering
\includegraphics[width=0.85\textwidth]{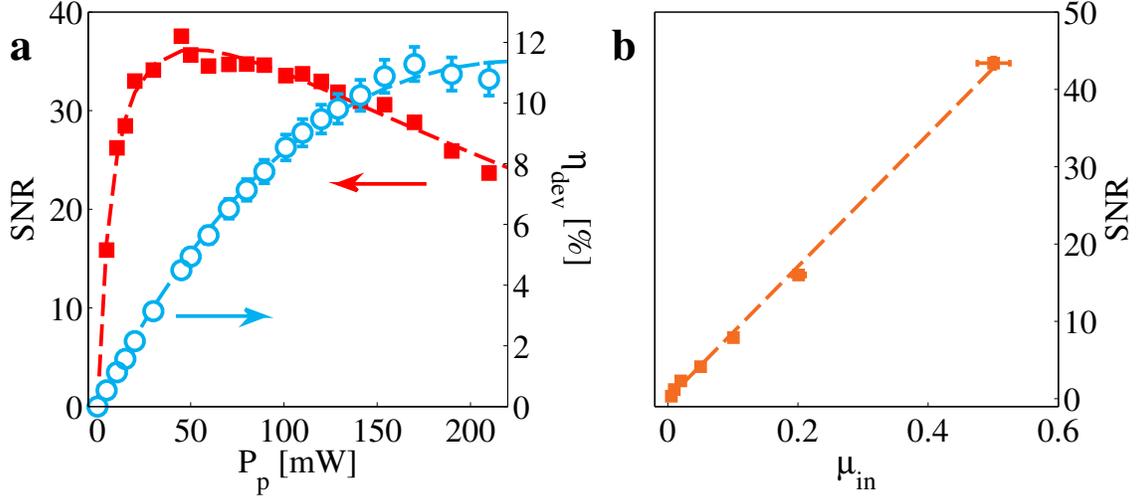}
\caption{{\bf QFCD characterization.} {\bf a}, Signal to noise
ratio (red plain squares, left axis) and device efficiency (blue
open circles, right axis) as a function of the pump power
(measured at the waveguide output) for $\mu_{\mathrm{in}} = 0.37$.
The dashed blue line corresponds to a fit of $\eta_{\mathrm{dev}}$
data using equation (\ref{eq:etatot}). The fit gives
$\eta_{\mathrm{max}} = 11.4(4)\,\%$ and
$\eta_{n}=119(9)\,\mathrm{\% W^{-1} cm^{-2}}$. The dashed red line
corresponds to a fit of the $\mathrm{SNR}$ data using equation
(\ref{eq:snr}).  {\bf b}, Signal to noise ratio as a function of
the mean input photon number $\mu_{\mathrm{in}}$ for $P_p \sim
120\,\mathrm{mW}$ (measured at the waveguide output). A linear fit
of the data (dashed orange line) gives
$\mathrm{SNR^{max}}=85(3)$.} \label{fig:S4}
\end{figure*}

In order to characterize the QFCD, we use weak coherent pulses at
$780\,\mathrm{nm}$ with a $\mathrm{FWHM}$ of $14\,\mathrm{ns}$, to
simulate the anti-Stokes photons obtained from our QM. We measure
the device efficiency and the signal to noise ratio for the
converted light as a function of the pump power. The results are
shown in Fig. \ref{fig:S4}a. The device efficiency varies as a
function of the pump power $P_{p}$ as:
\begin{equation}
\label{eq:etatot} \eta_{\mathrm{dev}} = \eta_{\mathrm{max}} \times
\sin^2 \left( L \sqrt{P_{p} \eta_{n}} \right),
\end{equation}
where $\eta_{\mathrm{max}}$ is the maximum achievable efficiency,
$L$ is the crystal length and $\eta_{n}$ is the normalized
conversion efficiency \cite{Zaske2011}. Fitting the experimental
data with equation (\ref{eq:etatot}), we find $\eta_{\mathrm{max}}
= 11.4(4)\,\%$ and $\eta_{n} = 119(9)\,\mathrm{\% W^{-1}
cm^{-2}}$. These values are compatible with the corresponding ones
obtained with quantum light at the device input and described in
the main paper. Despite the limited device efficiency measured in
the experiment, our QFCD already offers a significant advantage
over direct transmission of $780\,\mathrm{nm}$ photons over long
distances, as it is shown in Fig. \ref{fig:S2}.

An analogous formula can be derived for the signal to noise ratio
\cite{Fernandez-Gonzalvo2013}. We define the signal $s$ as
$\mu_{in} \cdot \eta_{\mathrm{dev}} \cdot \eta_{d,1552}$, where
$\mu_{in}$ is the mean input photon number per pulse and
$\eta_{d,1552}$ is the detection efficiency ($0.1$). The quantity
$s$ can be interpreted as the probability to get a click on the
detector due to a converted photon. On the other hand, the
probability to detect a noise photon is proportional to the pump
power, i.e. $n = \delta n \cdot P_{p}$. We therefore obtain:
\begin{equation}
\label{eq:snr} \mathrm{SNR} = \frac{s}{n + \mathrm{dc}},
\end{equation}
where $dc$ is the dark count probability. For low pump powers, the
device efficiency $\eta_{\mathrm{dev}}$ decreases (see equation
(\ref{eq:etatot})), thus causing a reduction of $s$. On the other
hand, the noise is limited to the dark count level $\mathrm{dc}$.
As a consequence, the $\mathrm{SNR}$ drops to zero. When $P_{p}$
increases, the $\mathrm{SNR}$ increases as well until a maximum is
reached. A further increase of the pump power will result in a
decrease of the $\mathrm{SNR}$, since the noise keeps on
increasing while the device efficiency saturates. This behavior
can clearly be observed in Fig. \ref{fig:S4}a. For the experiment
described in the main paper we used a pump power of about
$120\mathrm{mW}$ in order to operate the device close to the point
where the $\mathrm{SNR}$ is the highest.

As it will be detailed in the next section, the $\mathrm{SNR}$ is
the key parameter to estimate the $g^{(2)c}_{s,as}$ value
achievable after conversion. It is therefore important to have a
good estimate of its value. For this purpose, we measure the
$\mathrm{SNR}$ as a function of $\mu_{in}$ for a fixed pump power.
As it is shown in Fig. \ref{fig:S4}b, the $\mathrm{SNR}$ is
linearly proportional to $\mu_{in}$, i.e. $\mathrm{SNR} =
\mathrm{SNR^{max}} \times \mu_{in}$. The proportionality constant
$\mathrm{SNR^{max}}$ can be regarded as the maximum achievable
$\mathrm{SNR}$ when the system is operated with a single photon
input (i.e. $\mu_{in} = 1$). A fit to the experimental data give
us $\mathrm{SNR^{max}} = 85(3)$. This value is a factor $\sim70$
higher than what has been reported in
\cite{Fernandez-Gonzalvo2013}. This result has been achieved
thanks to the improvement of the noise filtering, the increase of
the device efficiency and the use of shorter photons.

\section{Expected second order cross-correlation function}
\label{app:C}
The second order cross-correlation function between the Stokes and
anti-Stokes fields experiences a reduction when the frequency of
the anti-Stokes photons is translated to the telecom band by means
of our QFCD, due to the presence of the broadband noise generated
by the pump. In this section we will detail the derivation of
equation 1 of the main paper, which allows us to quantify this
reduction.

The anti-Stokes detection probability after conversion can be
expressed by $p_{as}^{c} = \eta_{\mathrm{tot}} \cdot p_{as} +
p_n$, where $p_{as}$ is the probability to detect an unconverted
anti-Stokes photon, while $p_n$ is the probability to detect a
noise photon. The quantity $\eta_{\mathrm{tot}} =
\eta_{\mathrm{dev}} \eta_{\mathrm{loss}} \cdot (\eta_{d,1552} /
\eta_{d,780})$ accounts for the QFCD efficiency
($\eta_{\mathrm{dev}}$), the passive losses between the
anti-Stokes fiber output and the QFCD input
($\eta_{\mathrm{loss}}\sim0.77$) and the difference between the
efficiencies of the single photon detectors used for
$780\,\mathrm{nm}$ ($\eta_{d,780}=0.43$) and $1552\,\mathrm{nm}$
($\eta_{d,1552}=0.1$). The probability to detect an accidental
coincidence between the Stokes and anti-Stokes fields is defined
as $p_{as}^{c} \cdot p_s$ and it is therefore given by
$\eta_{\mathrm{tot}} \cdot p_s \cdot p_{as} + p_n \cdot p_s$. On
the other hand, the coincidence probability $p_{s,as}^{c}$ will be
given by $\eta_{\mathrm{tot}} \cdot p_{s,as} + p_n \cdot p_s$. The
factor $\eta_{\mathrm{tot}}$ is the same as for $p_{as}$, while
the quantity $p_n \cdot p_s$ can be regarded as the probability to
detect a noise photon conditioned on a previous detection of a
Stokes photon (conditional noise). Using the definition of the
second order cross-correlation function we obtain:
\begin{equation}
\label{eq:g2qfc1} g^{(2)c}_{s,as} = \frac{\eta_{\mathrm{tot}}
\cdot p_{s,as} + p_n \cdot p_s}{\eta_{\mathrm{tot}} \cdot p_s
\cdot p_{as} + p_n \cdot p_s}.
\end{equation}
We define the signal to noise ratio as $\mathrm{SNR} =
\eta_{\mathrm{tot}} \cdot p_{s,as} / (p_n \cdot p_s)$, which can
be interpreted as the signal to noise ratio for the detection of
an anti-Stokes photon conditioned on a previous detection of a
Stokes photon. With this definition equation (\ref{eq:g2qfc1}) can
be cast in the following form:
\begin{equation}
\label{eq:g2qfc2} g^{(2)c}_{s,as} = g^{(2)}_{s,as} \times
\frac{\mathrm{SNR} + 1}{\mathrm{SNR} + g^{(2)}_{s,as}}.
\end{equation}
This formula presents a few remarkable aspects. In the limit of
$g^{(2)}_{s,as}\rightarrow+\infty$ (as it happens for
$p_s\rightarrow0$), equation (\ref{eq:g2qfc2}) saturates to $1 +
\mathrm{SNR}$. The signal to noise ratio therefore expresses the
maximum cross-correlation level achievable after conversion. On
the other hand, for high Stokes detection probabilities the
$g^{(2)}_{s,as}$ function decreases. If we assume that
$\mathrm{SNR}\gg1$, then equation (\ref{eq:g2qfc2}) gives
$g^{(2)c}_{s,as} \sim g^{(2)}_{s,as}$.

\begin{figure*}[hbt]
\centering
\includegraphics[width=0.85\textwidth]{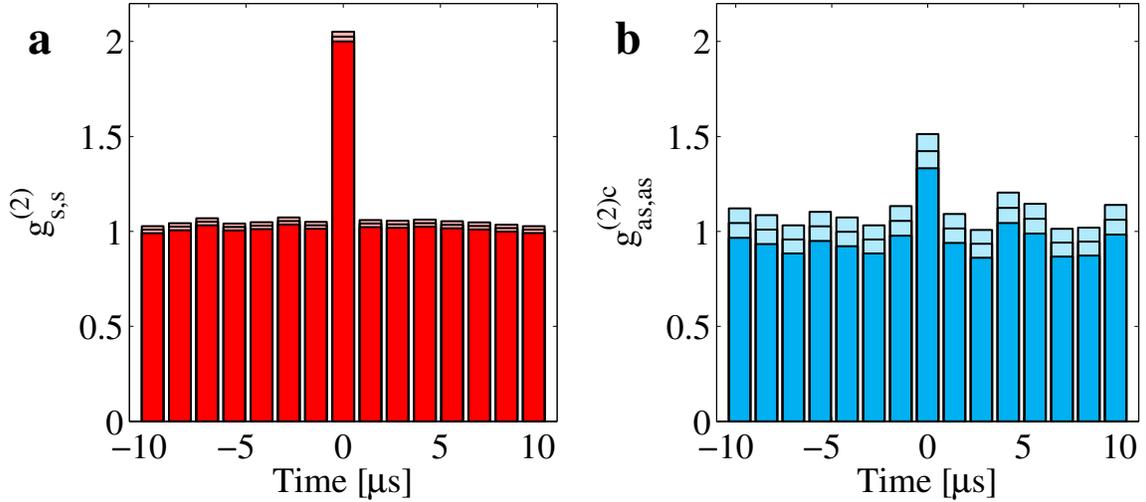}
\caption{{\bf Autocorrelation measurements.} {\bf a},
Autocorrelation measurement for the Stokes field. The histogram
has been obtained combining the data corresponding to different
Stokes detection probabilities. The measured $g^{(2)}_{s,s}$ value
is $2.03(3)$ (see table \ref{tab:g2auto}). {\bf b},
Autocorrelation measurement for the converted anti-Stokes photons.
The histogram is obtained combining all the data shown in table
\ref{tab:g2auto}. The measured $g^{(2)c}_{as,as}$ value is
$1.4(1)$.} \label{fig:S5}
\end{figure*}

As we discussed in the previous section, the $\mathrm{SNR}$
depends linearly on the probability to have a photon at the input
of the QFCD. When we convert heralded single photons from the QM,
this probability is given by the retrieval efficiency
$\eta_R^{in}$. Since $\eta_R^{in}$ is constant over a wide range
of Stokes detection probabilities (see Fig. 2 of the main text),
for a fixed pump power the $\mathrm{SNR}$ can also be taken as
constant in the same $p_s$ range. This assumption does not hold
for high $p_s$ values. However, as we discussed above, in this
case equation (\ref{eq:g2qfc2}) does not depend on the
$\mathrm{SNR}$. Therefore the $\mathrm{SNR}$ can be assumed as
constant.

\section{Cauchy-Schwarz inequality and autocorrelation
measurements}\label{app:D} For a pair of independent classical
fields, the second order cross-correlation function between the
two modes and the relative autocorrelation functions for each mode
have to satisfy the following inequality:
\begin{equation}
\label{eq:cauchy} R =
\frac{\left(g^{(2)}_{1,2}\right)^2}{g^{(2)}_{1,1} \cdot
g^{(2)}_{2,2}} \leq 1,
\end{equation}
which constitutes a particular case of the Cauchy-Schwarz
inequality \cite{Clauser1974}. A violation of the condition
expressed by equation (\ref{eq:cauchy}) indicates that
non-classical correlations are present between the two modes. As
we explained before, the joint quantum state for the Stokes and
anti-Stokes field can be described by the two mode squeezed state
defined in equation (\ref{eq:squeezed}). In the ideal case, the
Stokes and the anti-Stokes modes individually display a thermal
statistic, i.e. their second order autocorrelation is equal to
$2$. Using equation (\ref{eq:cauchy}), we see that when the
cross-correlation function $g^{(2)}_{s,as}$ exceeds the value of
$2$, the joint photon statistic shows a quantum behavior. However,
for a generic two modes state, the autocorrelation function of
each mode can exceed the value of $2$. In this case, it is
necessary to achieve higher $g^{(2)}_{s,as}$ values in order to
ensure that the Cauchy-Schwarz inequality is violated. For this
reason, it is necessary to carefully measure the Stokes and
anti-Stokes autocorrelations before assessing the quantum nature
of the observed state without assumptions.

In order to measure the Stokes autocorrelation function, we use a
$50$-$50$ fiber beam splitter to deliver the photons obtained from
our QM to two single photon detectors. The arrival times on the
two detectors are measured by means of the time stamping card
mentioned above, thus obtaining a coincidence histogram. The
$g^{(2)}_{s,s}$ is shown in Fig. \ref{fig:S5}a. We proceed in a
similar way for the anti-Stokes field. The measurements are
performed for five different {\it write} powers. The results are
summarized in the first three columns of table \ref{tab:g2auto},
where the two autocorrelation values are listed together with the
corresponding measured Stokes detection probabilities. We can
observe that the autocorrelation values do not change
significantly for the $p_s$ range explored in the experiment. For
this reason, we can combine all the histograms together and obtain
a unique value for the autocorrelations, shown in the last line of
the table. The measured $g^{(2)}_{s,s}$ is compatible with the
expected thermal statistic. For the anti-Stokes field the observed
value is systematically below $2$. This can be explained
considering that for the anti-Stokes photons we do not use
spectral filtering to further reduce the possible leakage of {\it
read} light. For this reason the signal to noise ratio is lower in
this case. In the fifth column of table \ref{tab:g2auto} we show
the value of the Cauchy-Schwarz parameter $R$ calculated from the
$g^{(2)}_{s,as}$ measured in the experiment.

\begin{table}
\begin{center}
\caption{Values of the Stokes ($g^{(2)}_{s,s}$), anti-Stokes
($g^{(2)}_{as,as}$) and converted anti-Stokes ($g^{(2)c}_{as,as}$)
autocorrelation function for various Stokes detection
probabilities $p_s$. The values shown in the last row are obtained
combining all the correlation histograms. In the last two columns
the Cauchy-Schwarz parameter $R$ ($R^c$) is calculated from the
measured $g^{(2)}_{s,as}$ ($g^{(2)c}_{s,as}$) values.}
\label{tab:g2auto}
\begin{tabular}{c|c|c|c||c|c}
$p_s$ [$\%$] & $g^{(2)}_{s,s}$ & $g^{(2)}_{as,as}$ & $g^{(2)c}_{as,as}$ & $R$ & $R^{c}$\\
\hline
$1.6$  & $2.03(5)$ & $1.46(2)$ & $1.6(2)$ & $12.2(4)$ & $4.8(6)$ \\
$1.5$  & $2.01(6)$ & $1.46(2)$ & $1.3(2)$ & $12.8(5)$ & $6(1)$   \\
$1.0$  & $2.03(8)$ & $1.41(2)$ & $1.4(2)$ & $16.4(8)$ & $7(1)$   \\
$0.4$  & $2.03(5)$ & $1.43(4)$ & $1.3(2)$ & $55(2)$   & $21(5)$  \\
\hline
       & $2.03(3)$ & 1.45(1) & 1.4(1) \\
\end{tabular}
\end{center}
\end{table}

It is possible to proceed in a similar way when the atomic based
QM is used in combination with the QFCD. For the Stokes field, we
can use the same value as before. For the anti-Stokes field, a new
measurement is necessary, performed following the same
experimental strategy as above (see Fig. \ref{fig:S5}b). The
results are shown in the forth column of table \ref{tab:g2auto},
together with their average. The data have been taken using
$120\,\mathrm{mW}$ of pump power, measured at the waveguide
output. We notice that the measured $g^{(2)c}_{as,as}$ value is
below $2$. Similar reasons as before contribute to explain this
behavior: the converted anti-Stokes photons are polluted by the
broadband noise generated by the strong pump at
$1569\,\mathrm{nm}$ while propagating in the non linear waveguide.
The integration time used to accumulate the autocorrelation
histogram corresponding the lowest $p_s$ value reported in table
\ref{tab:g2auto} is about $10$ hours. The Cauchy-Schwarz parameter
$R^c$ shown in the last column is clearly above $1$, thus
confirming that the quantum nature of the anti-Stokes photons is
preserved in the conversion process.

\begin{acknowledgments}
The authors acknowledge financial support by the ERC starting
grant QuLIMA, by the Spanish MINECO (project FIS2012-37569) and by
the US Army RDECOM. We thank Giacomo Corrielli and Marcel$\cdot$li
Grimau for help during the early stage of the experiment.
\end{acknowledgments}


\begin{thebibliography}{10}
\expandafter\ifx\csname url\endcsname\relax
  \def\url#1{\texttt{#1}}\fi
\expandafter\ifx\csname
urlprefix\endcsname\relax\def\urlprefix{URL }\fi
\providecommand{\bibinfo}[2]{#2}
\providecommand{\eprint}[2][]{\url{#2}}

\bibitem{Kimble2008}
\bibinfo{author}{Kimble, H.~J.}
\newblock \bibinfo{title}{The quantum internet}.
\newblock \emph{\bibinfo{journal}{Nature}} \textbf{\bibinfo{volume}{453}},
  \bibinfo{pages}{1023--1030} (\bibinfo{year}{2008}).

\bibitem{Hammerer2010}
\bibinfo{author}{Hammerer, K.}, \bibinfo{author}{S{\o}rensen, A.~S.} \&
  \bibinfo{author}{Polzik, E.~S.}
\newblock \bibinfo{title}{Quantum interface between light and atomic
  ensembles}.
\newblock \emph{\bibinfo{journal}{Rev. Mod. Phys.}}
  \textbf{\bibinfo{volume}{82}}, \bibinfo{pages}{1041--1093}
  (\bibinfo{year}{2010}).

\bibitem{Sangouard2011}
\bibinfo{author}{Sangouard, N.}, \bibinfo{author}{Simon, C.},
  \bibinfo{author}{de~Riedmatten, H.} \& \bibinfo{author}{Gisin, N.}
\newblock \bibinfo{title}{Quantum repeaters based on atomic ensembles and
  linear optics}.
\newblock \emph{\bibinfo{journal}{Rev. Mod. Phys.}}
  \textbf{\bibinfo{volume}{83}}, \bibinfo{pages}{33--80}
  (\bibinfo{year}{2011}).

\bibitem{Ritter2012}
\bibinfo{author}{Ritter, S.} \emph{et~al.}
\newblock \bibinfo{title}{An elementary quantum network of single atoms in
  optical cavities}.
\newblock \emph{\bibinfo{journal}{Nature}} \textbf{\bibinfo{volume}{484}},
  \bibinfo{pages}{195--200} (\bibinfo{year}{2012}).

\bibitem{Moehring2007}
\bibinfo{author}{Moehring, D.~L.} \emph{et~al.}
\newblock \bibinfo{title}{Entanglement of single-atom quantum bits at a
  distance}.
\newblock \emph{\bibinfo{journal}{Nature}} \textbf{\bibinfo{volume}{449}},
  \bibinfo{pages}{68--71} (\bibinfo{year}{2007}).

\bibitem{Hofmann2012}
\bibinfo{author}{Hofmann, J.} \emph{et~al.}
\newblock \bibinfo{title}{Heralded entanglement between widely separated
  atoms}.
\newblock \emph{\bibinfo{journal}{Science}} \textbf{\bibinfo{volume}{337}},
  \bibinfo{pages}{72--75} (\bibinfo{year}{2012}).

\bibitem{Chaneliere2005}
\bibinfo{author}{Chaneli\`{e}re, T.} \emph{et~al.}
\newblock \bibinfo{title}{Storage and retrieval of single photons transmitted
  between remote quantum memories}.
\newblock \emph{\bibinfo{journal}{Nature}} \textbf{\bibinfo{volume}{438}},
  \bibinfo{pages}{833--836} (\bibinfo{year}{2005}).

\bibitem{Eisaman2005}
\bibinfo{author}{Eisaman, M.~D.} \emph{et~al.}
\newblock \bibinfo{title}{Electromagnetically induced transparency with tunable
  single-photon pulses}.
\newblock \emph{\bibinfo{journal}{Nature}} \textbf{\bibinfo{volume}{438}},
  \bibinfo{pages}{837--841} (\bibinfo{year}{2005}).

\bibitem{Radnaev2010}
\bibinfo{author}{Radnaev, A.~G.} \emph{et~al.}
\newblock \bibinfo{title}{A quantum memory with telecom-wavelength conversion}.
\newblock \emph{\bibinfo{journal}{Nat Phys}} \textbf{\bibinfo{volume}{6}},
  \bibinfo{pages}{894--899} (\bibinfo{year}{2010}).

\bibitem{Hosseini2011}
\bibinfo{author}{Hosseini, M.}, \bibinfo{author}{Campbell, G.},
  \bibinfo{author}{Sparkes, B.~M.}, \bibinfo{author}{Lam, P.~K.} \&
  \bibinfo{author}{Buchler, B.~C.}
\newblock \bibinfo{title}{Unconditional room-temperature quantum memory}.
\newblock \emph{\bibinfo{journal}{Nat Phys}} \textbf{\bibinfo{volume}{7}},
  \bibinfo{pages}{794--798} (\bibinfo{year}{2011}).

\bibitem{Bao2012}
\bibinfo{author}{Bao, X.-H.} \emph{et~al.}
\newblock \bibinfo{title}{Efficient and long-lived quantum memory with cold
  atoms inside a ring cavity}.
\newblock \emph{\bibinfo{journal}{Nat Phys}} \textbf{\bibinfo{volume}{8}},
  \bibinfo{pages}{517--521} (\bibinfo{year}{2012}).

\bibitem{Riedmatten2008}
\bibinfo{author}{de~Riedmatten, H.}, \bibinfo{author}{Afzelius, M.},
  \bibinfo{author}{Staudt, M.~U.}, \bibinfo{author}{Simon, C.} \&
  \bibinfo{author}{Gisin, N.}
\newblock \bibinfo{title}{A solid-state light-matter interface at the
  single-photon level}.
\newblock \emph{\bibinfo{journal}{Nature}} \textbf{\bibinfo{volume}{456}},
  \bibinfo{pages}{773--777} (\bibinfo{year}{2008}).

\bibitem{Hedges2010}
\bibinfo{author}{Hedges, M.~P.}, \bibinfo{author}{Longdell, J.~J.},
  \bibinfo{author}{Li, Y.} \& \bibinfo{author}{Sellars, M.~J.}
\newblock \bibinfo{title}{Efficient quantum memory for light}.
\newblock \emph{\bibinfo{journal}{Nature}} \textbf{\bibinfo{volume}{465}},
  \bibinfo{pages}{1052--1056} (\bibinfo{year}{2010}).

\bibitem{Clausen2011}
\bibinfo{author}{Clausen, C.} \emph{et~al.}
\newblock \bibinfo{title}{Quantum storage of photonic entanglement in a
  crystal}.
\newblock \emph{\bibinfo{journal}{Nature}} \textbf{\bibinfo{volume}{469}},
  \bibinfo{pages}{508--511} (\bibinfo{year}{2011}).

\bibitem{Saglamyurek2011}
\bibinfo{author}{Saglamyurek, E.} \emph{et~al.}
\newblock \bibinfo{title}{Broadband waveguide quantum memory for entangled
  photons}.
\newblock \emph{\bibinfo{journal}{Nature}} \textbf{\bibinfo{volume}{469}},
  \bibinfo{pages}{512--515} (\bibinfo{year}{2011}).

\bibitem{Chou2007}
\bibinfo{author}{Chou, C.~W.} \emph{et~al.}
\newblock \bibinfo{title}{Functional quantum nodes for entanglement
  distribution over scalable quantum networks}.
\newblock \emph{\bibinfo{journal}{Science}} \textbf{\bibinfo{volume}{316}},
  \bibinfo{pages}{1316--1320} (\bibinfo{year}{2007}).
\newblock

\bibitem{Yuan2008}
\bibinfo{author}{Yuan, Z.-S.} \emph{et~al.}
\newblock \bibinfo{title}{Experimental demonstration of a bdcz quantum repeater
  node}.
\newblock \emph{\bibinfo{journal}{Nature}} \textbf{\bibinfo{volume}{454}},
  \bibinfo{pages}{1098--1101} (\bibinfo{year}{2008}).

\bibitem{Usmani2012}
\bibinfo{author}{Usmani, I.} \emph{et~al.}
\newblock \bibinfo{title}{Heralded quantum entanglement between two crystals}.
\newblock \emph{\bibinfo{journal}{Nat Photon}} \textbf{\bibinfo{volume}{6}},
  \bibinfo{pages}{234--237} (\bibinfo{year}{2012}).

\bibitem{Briegel1998}
\bibinfo{author}{Briegel, H.-J.}, \bibinfo{author}{D\"ur, W.},
  \bibinfo{author}{Cirac, J.~I.} \& \bibinfo{author}{Zoller, P.}
\newblock \bibinfo{title}{Quantum repeaters: The role of imperfect local
  operations in quantum communication}.
\newblock \emph{\bibinfo{journal}{Phys. Rev. Lett.}}
  \textbf{\bibinfo{volume}{81}}, \bibinfo{pages}{5932--5935}
  (\bibinfo{year}{1998}).

\bibitem{Duan2001}
\bibinfo{author}{Duan, L.-M.}, \bibinfo{author}{Lukin, M.~D.},
  \bibinfo{author}{Cirac, J.~I.} \& \bibinfo{author}{Zoller, P.}
\newblock \bibinfo{title}{Long-distance quantum communication with atomic
  ensembles and linear optics}.
\newblock \emph{\bibinfo{journal}{Nature}} \textbf{\bibinfo{volume}{414}},
  \bibinfo{pages}{413--418} (\bibinfo{year}{2001}).

\bibitem{Lauritzen2010}
\bibinfo{author}{Lauritzen, B.} \emph{et~al.}
\newblock \bibinfo{title}{Telecommunication-wavelength solid-state memory at
  the single photon level}.
\newblock \emph{\bibinfo{journal}{Phys. Rev. Lett.}}
  \textbf{\bibinfo{volume}{104}}, \bibinfo{pages}{080502--}
  (\bibinfo{year}{2010}).
\newblock

\bibitem{Tanzilli2005}
\bibinfo{author}{Tanzilli, S.} \emph{et~al.}
\newblock \bibinfo{title}{A photonic quantum information interface}.
\newblock \emph{\bibinfo{journal}{Nature}} \textbf{\bibinfo{volume}{437}},
  \bibinfo{pages}{116--120} (\bibinfo{year}{2005}).

\bibitem{Rakher2010}
\bibinfo{author}{Rakher, M.~T.}, \bibinfo{author}{Ma, L.},
  \bibinfo{author}{Slattery, O.}, \bibinfo{author}{Tang, X.} \&
  \bibinfo{author}{Srinivasan, K.}
\newblock \bibinfo{title}{Quantum transduction of telecommunications-band
  single photons from a quantum dot by frequency upconversion}.
\newblock \emph{\bibinfo{journal}{Nat Photon}} \textbf{\bibinfo{volume}{4}},
  \bibinfo{pages}{786--791} (\bibinfo{year}{2010}).

\bibitem{Ates2012}
\bibinfo{author}{Ates, S.} \emph{et~al.}
\newblock \bibinfo{title}{Two-photon interference using background-free quantum
  frequency conversion of single photons emitted by an inas quantum dot}.
\newblock \emph{\bibinfo{journal}{Phys. Rev. Lett.}}
  \textbf{\bibinfo{volume}{109}}, \bibinfo{pages}{147405--}
  (\bibinfo{year}{2012}).
\newblock

\bibitem{Takesue2010}
\bibinfo{author}{Takesue, H.}
\newblock \bibinfo{title}{Single-photon frequency down-conversion experiment}.
\newblock \emph{\bibinfo{journal}{Phys. Rev. A}} \textbf{\bibinfo{volume}{82}},
  \bibinfo{pages}{013833--} (\bibinfo{year}{2010}).

\bibitem{Curtz2010}
\bibinfo{author}{Curtz, N.}, \bibinfo{author}{Thew, R.},
  \bibinfo{author}{Simon, C.}, \bibinfo{author}{Gisin, N.} \&
  \bibinfo{author}{Zbinden, H.}
\newblock \bibinfo{title}{Coherent frequency-down-conversion interface for
  quantum repeaters}.
\newblock \emph{\bibinfo{journal}{Opt. Express}} \textbf{\bibinfo{volume}{18}},
  \bibinfo{pages}{22099--22104} (\bibinfo{year}{2010}).
\newblock

\bibitem{Zaske2011}
\bibinfo{author}{Zaske, S.}, \bibinfo{author}{Lenhard, A.} \&
  \bibinfo{author}{Becher, C.}
\newblock \bibinfo{title}{Efficient frequency downconversion at the single
  photon level from the red spectral range to the telecommunications c-band}.
\newblock \emph{\bibinfo{journal}{Opt. Express}} \textbf{\bibinfo{volume}{19}},
  \bibinfo{pages}{12825--12836} (\bibinfo{year}{2011}).
\newblock

\bibitem{Zaske2012}
\bibinfo{author}{Zaske, S.} \emph{et~al.}
\newblock \bibinfo{title}{Visible-to-telecom quantum frequency conversion of
  light from a single quantum emitter}.
\newblock \emph{\bibinfo{journal}{Phys. Rev. Lett.}}
  \textbf{\bibinfo{volume}{109}}, \bibinfo{pages}{147404--}
  (\bibinfo{year}{2012}).
\newblock

\bibitem{Ikuta2011}
\bibinfo{author}{Ikuta, R.} \emph{et~al.}
\newblock \bibinfo{title}{Wide-band quantum interface for
  visible-to-telecommunication wavelength conversion}.
\newblock \emph{\bibinfo{journal}{Nat Commun}} \textbf{\bibinfo{volume}{2}},
  \bibinfo{pages}{537--} (\bibinfo{year}{2011}).

\bibitem{DeGreve2012}
\bibinfo{author}{De~Greve, K.} \emph{et~al.}
\newblock \bibinfo{title}{Quantum-dot spin-photon entanglement via frequency
  downconversion to telecom wavelength}.
\newblock \emph{\bibinfo{journal}{Nature}} \textbf{\bibinfo{volume}{491}},
  \bibinfo{pages}{421--425} (\bibinfo{year}{2012}).

\bibitem{Felinto2006}
\bibinfo{author}{Felinto, D.} \emph{et~al.}
\newblock \bibinfo{title}{Conditional control of the quantum states of remote
  atomic memories for quantum networking}.
\newblock \emph{\bibinfo{journal}{Nat Phys}} \textbf{\bibinfo{volume}{2}},
  \bibinfo{pages}{844--848} (\bibinfo{year}{2006}).

\bibitem{Laurat2006}
\bibinfo{author}{Laurat, J.} \emph{et~al.}
\newblock \bibinfo{title}{Efficient retrieval of a single excitation stored in
  an atomic ensemble}.
\newblock \emph{\bibinfo{journal}{Opt. Express}} \textbf{\bibinfo{volume}{14}},
  \bibinfo{pages}{6912--6918} (\bibinfo{year}{2006}).
\newblock

\bibitem{Fernandez-Gonzalvo2013}
\bibinfo{author}{Fernandez-Gonzalvo, X.} \emph{et~al.}
\newblock \bibinfo{title}{Quantum frequency conversion of quantum memory
  compatible photons to telecommunication wavelengths}.
\newblock \emph{\bibinfo{journal}{Opt. Express}} \textbf{\bibinfo{volume}{21}},
  \bibinfo{pages}{19473--19487} (\bibinfo{year}{2013}).
\newblock

\bibitem{Riedmatten2006}
\bibinfo{author}{de~Riedmatten, H.} \emph{et~al.}
\newblock \bibinfo{title}{Direct measurement of decoherence for entanglement
  between a photon and stored atomic excitation}.
\newblock \emph{\bibinfo{journal}{Phys. Rev. Lett.}}
  \textbf{\bibinfo{volume}{97}}, \bibinfo{pages}{113603--4}
  (\bibinfo{year}{2006}).

\bibitem{Sangouard2008a}
\bibinfo{author}{Sangouard, N.} \emph{et~al.}
\newblock \bibinfo{title}{Robust and efficient quantum repeaters with atomic
  ensembles and linear optics}.
\newblock \emph{\bibinfo{journal}{Phys.Rev.A}} \textbf{\bibinfo{volume}{77}},
  \bibinfo{pages}{062301} (\bibinfo{year}{2008}).

\bibitem{Palittapongarnpim2012}
\bibinfo{author}{Palittapongarnpim, P.}, \bibinfo{author}{MacRae, A.} \&
  \bibinfo{author}{Lvovsky, A.~I.}
\newblock \bibinfo{title}{Note: A monolithic filter cavity for experiments in
  quantum optics}.
\newblock \emph{\bibinfo{journal}{Rev. Sci. Instrum.}}
  \textbf{\bibinfo{volume}{83}}, \bibinfo{pages}{066101--3}
  (\bibinfo{year}{2012}).

\bibitem{Mendes2013}
\bibinfo{author}{Mendes, M.~S.}, \bibinfo{author}{Saldanha, P.~L.},
  \bibinfo{author}{Tabosa, J. W.~R.} \& \bibinfo{author}{Felinto, D.}
\newblock \bibinfo{title}{Dynamics of the reading process of a quantum memory}.
\newblock \emph{\bibinfo{journal}{New Journal of Physics}}
  \textbf{\bibinfo{volume}{15}}, \bibinfo{pages}{075030--}
  (\bibinfo{year}{2013}).

\bibitem{kuo2013}
\bibinfo{author}{Kuo, P.~S.} \emph{et~al.}
\newblock \bibinfo{title}{Reducing noise in single-photon-level frequency
  conversion}.
\newblock \emph{\bibinfo{journal}{Opt. Lett.}} \textbf{\bibinfo{volume}{38}},
  \bibinfo{pages}{1310--1312} (\bibinfo{year}{2013}).

\bibitem{Clauser1974}
\bibinfo{author}{Clauser, J.~F.}
\newblock \bibinfo{title}{Experimental distinction between the quantum and
  classical field-theoretic predictions for the photoelectric effect}.
\newblock \emph{\bibinfo{journal}{Phys. Rev. D}} \textbf{\bibinfo{volume}{9}},
  \bibinfo{pages}{853--} (\bibinfo{year}{1974}).

\end{thebibliography}
\end{document}